\documentclass[aps,reprint,onecolumn,showpacs,superscriptaddress]{revtex4-1}

\usepackage{graphicx}
\usepackage{dcolumn}
\usepackage{bm}
\usepackage{amssymb,amsfonts,amsmath}
\usepackage{color}

\newcommand\ie{i.e.\ }

\begin{document}

\title{Confined Rayleigh-B\'enard, Rotating Rayleigh-B\'enard, and Double Diffusive Convection: A unifying view on turbulent transport enhancement through coherent structure manipulation}

\author{Kai Leong Chong}
\affiliation{Department of Physics, The Chinese University of Hong Kong, Shatin, Hong Kong, China}
\author{Yantao Yang}
\affiliation{Physics of Fluids Group, MESA+ Institute and J. M. Burgers Centre for Fluid Dynamics, University of Twente, 7500 AE Enschede, The Netherlands}
\author{Shi-Di Huang}
\affiliation{Department of Physics, The Chinese University of Hong Kong, Shatin, Hong Kong, China}
\author{Jin-Qiang Zhong}
\affiliation{Shanghai Key Laboratory of Special Artificial Microstructure Materials and Technology and School of Physics Science and Engineering, Tongji University, Shanghai 200092, China}
\author{Richard J.A.M. Stevens}
\affiliation{Physics of Fluids Group, MESA+ Institute and J. M. Burgers Centre for Fluid Dynamics, University of Twente, 7500 AE Enschede, The Netherlands}
\author{Roberto Verzicco}
\affiliation{Physics of Fluids Group, MESA+ Institute and J. M. Burgers Centre for Fluid Dynamics, University of Twente, 7500 AE Enschede, The Netherlands}
\affiliation{Dipartimento di Ingegneria Industriale, University of Rome Tor Vergata, Rome 00133, Italy}
\author{Detlef Lohse}
\affiliation{Physics of Fluids Group, MESA+ Institute and J. M. Burgers Centre for Fluid Dynamics, University of Twente, 7500 AE Enschede, The Netherlands}
\affiliation{Max-Planck Institute for Dynamics and Self-Organization, 37077 G\"ottingen, Germany}
\author{Ke-Qing Xia}
\email{kxia@cuhk.edu.hk}
\affiliation{Department of Physics, The Chinese University of Hong Kong, Shatin, Hong Kong, China}
\date{\today}

\begin{abstract}
Many natural and engineering systems are simultaneously subjected to a driving force and a stabilizing force. The interplay between the two forces, especially for highly nonlinear systems such as fluid flow, often results in surprising features. Here we reveal such features in three different types of Rayleigh-B\'enard (RB) convection, i.e. buoyancy-driven flow with the fluid density being affected by a scalar field. In the three cases different {\it stabilizing forces} are considered, namely (i) horizontal confinement, (ii) rotation around a vertical axis, and (iii) a second stabilizing scalar field. Despite the very different nature of the stabilizing forces and the corresponding equations of motion, at moderate strength we counterintuitively but consistently observe an {\it enhancement} in the flux, even though the flow motion is weaker than the original RB flow. The flux enhancement occurs in an intermediate regime in which the stabilizing force is strong enough to alter the flow structures in the bulk to a more organised morphology, yet not too strong to severely suppress the flow motions. Near the optimal transport enhancements all three systems exhibit a transition from a state in which the thermal boundary layer (BL)  is nested inside the momentum BL to the one with the thermal BL being thicker than the momentum BL. 
\end{abstract}


\maketitle

It is very common in nature and engineering settings that, in addition to a driving force, a system is also subjected to a  stabilizing force. For a highly nonlinear system, the presence of the stabilizing force may induce surprising phenomena. For instance, Rayleigh-B\'enard (RB) convection, which is in nature commonly encountered~\citep{ahlers2009rmp,lohse2010arfm,chilla2012epj,xia2013taml} buoyancy driven unstably stratified flow, often experiences a {\it stabilizing} mechanism. The first example is RB convection under lateral geometrical confinement (CRB). Here the buoyancy driving interplays with the viscous force from sidewalls. The second example is RB convection under rotation (RRB) in which the Coriolis force is well-known to have a stabilizing effect that can be understood in terms of the Taylor-Proudman theorem \citep{proudman1916prsla,taylor1917prsla,chandrasekhar2013hydrodynamic}. Our third example is double diffusive convection (DDC)~\citep{radko2013double}, where the fluid density is determined by two scalars with different molecular diffusivities, such as temperature and salinity in seawater. Here the two scalars can have opposite stratification, leading to a stabilization force for the unstably stratified scalar field. All these three systems are of great importance in astrophysics \citep{busse1994chaos,merryfield1995aj,rosenblum2011aj,leconte2012aa,mirouh2012aj,wood2013aj}, geophysics \citep{schoofs1999epsl,buffett2010jgrse}, oceanography \citep{turner1985arfm,schmitt1994arfm,marshall1999rg,schmitt2005science} and engineering applications \citep{nada2007ijhmt}. 

In these three systems, the stabilizing forces are completely different and correspondingly different physical parameters are  required to quantify the degree of stabilization. In CRB, it is the reciprocal of the width-to-height ratio $1/\Gamma$ that characterizes the relative strength of stabilizing~\citep{huang2013prl,wagner2013pof,chong2015prl}. In RRB, the stabilization is characterized by the ratio of Coriolis force to buoyancy which is the reciprocal Rossby number $1/Ro$~\citep{king2009nature,zhong2009prl,stevens2009prl,wei2015prl}. In DDC, it is the ratio of the buoyancy force induced by temperature gradient to that by the salinity gradient, \ie the density ratio $\Lambda$~\citep{kellner2014pof,yang2016pnas}, that characterizes the relative strength of stabilization. Since the stabilizing mechanisms in the three systems are very different, one would expect that CRB, RRB and DDC will behave very differently when subjected to the respective stabilizing forces. In this Letter, however, we will show that the salient features in the three seemingly different systems are universal and can all be explained by coherent structure manipulation and boundary layer crossing and therefore can be understood in terms of a unifying framework.

\begin{figure*}[b!]
\includegraphics[width=1.0\textwidth]{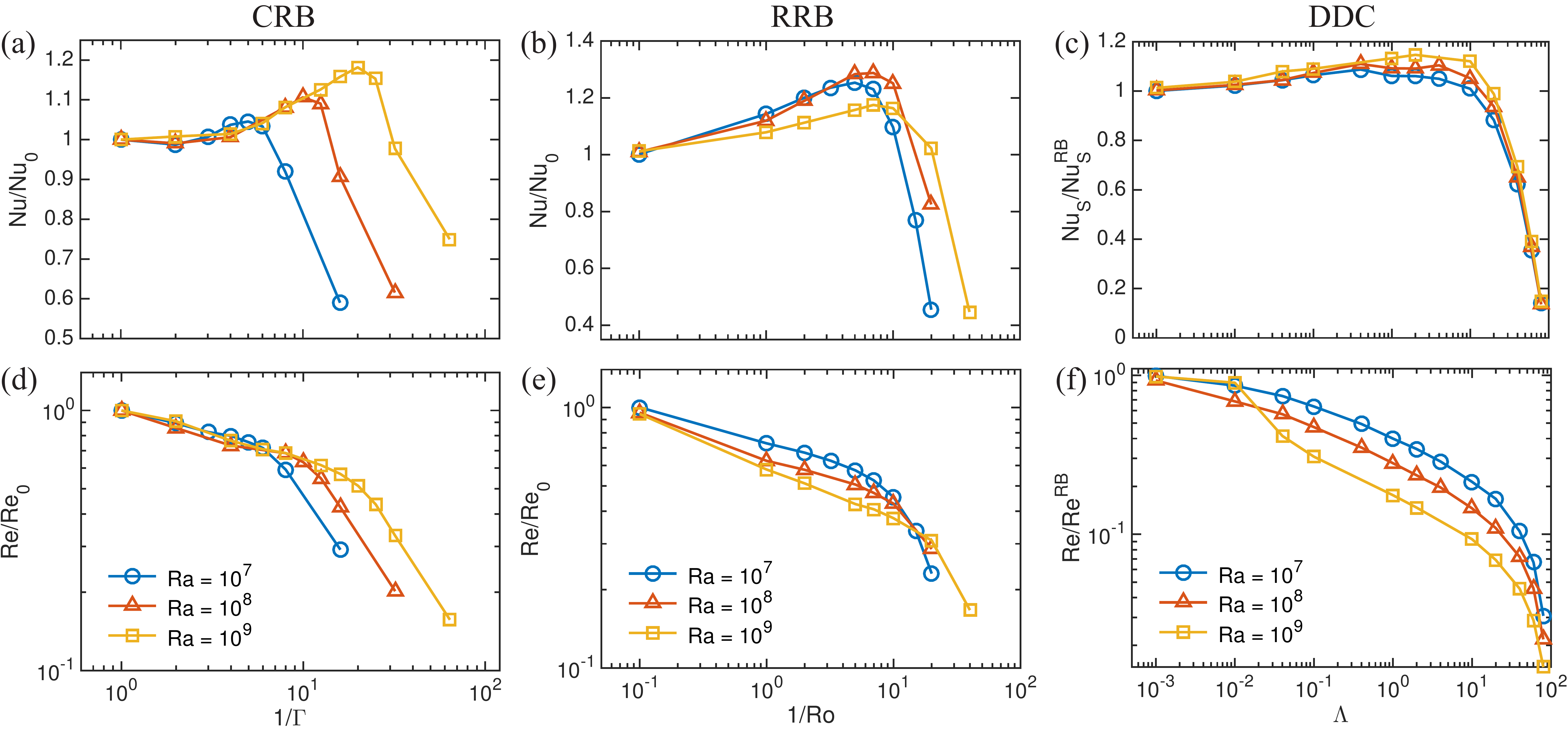}
\caption{\label{fig:NuRe}Nusselt number $Nu$ or $Nu_S$ and Reynolds number $Re$ versus $1/\Gamma$ for CRB in (a) and (d), $1/Ro$ for RRB in (b) and (e) and $\Lambda$ for DDC in (c) and (f). Both quantities are normalized by the value obtained from cases $1/\Gamma=1$, $1/Ro=0$ or $\Lambda=0$ (represented by $Nu_0$ and $Re_0$ in CRB and RRB while represented by $Nu_S^{RB}$ and $Re^{RB}$ in DDC).}
\end{figure*}

Our analysis is based on datasets obtained from direct numerical simulations (DNS) of the three systems. For the simulations, the incompressible Navier-Stokes equation within the Oberbeck-Boussinesq approximations and the convection-diffusion equation(s) are solved for velocities and the scalar field(s), where the Coriolis force and the additional buoyancy gradient generated by a stable temperature gradient are added for RRB and DDC, respectively. The physical quantities are nondimensionalized by the cell height $H$, the global temperature/salinity difference $\Delta_T/\Delta_S$ and the free-fall velocity. The data are taken from our previous simulations reported in Refs.~\cite{kaczorowski2008nrnefm6,kaczorowski2013jfm,kaczorowski2014jfm} for CRB and Refs.~\cite{yang2015jfm,yang2016pnas,yang2016jfm} for DDC, respectively. The RRB simulations were conducted by using the numerical solver described in Refs.~\cite{ostillamonico2015jcp}. For each system, three Rayleigh numbers are presented, which are $Ra=10^7$, $10^8$ and $10^9$. For each fixed $Ra$, simulations were conducted for a wide range of $1/\Gamma$, $1/Ro$ and $\Lambda$. The Prandtl number is $Pr=4.38$ in CRB, $Pr=6.4$ in RRB and the Lewis number in DDC is $Le=100$. In RRB and DDC periodic boundary conditions are applied in the horizontal directions and the box size is set to be much larger than the horizontal width of typical flow structures.

In Fig.~\ref{fig:NuRe} we show the Nusselt number $Nu$ and Reynolds number $Re$ versus the degree of stabilization in the three systems. In CRB and RRB, the heat transport is considered while in DDC the transport of the primary scalar, i.e. salinity, is considered instead. For the three systems, the Reynolds number is evaluated based on the root-mean-square velocity averaged over the whole domain and over time. As the strengths of stabilization increase, the global transport behaviour undergoes a transition from a typical RB regime to the regime dominated by the stabilizing force. However, earlier studies in CRB \citep{huang2013prl,chong2015prl}, RRB \cite{zhong2009prl,stevens2009prl} and DDC \cite{yang2016pnas,yang2016jfm} separately revealed an intermediate regime with enhancement in scalar transport, and in CRB and DDC even the decoupling of scalar and mass transport was observed. Figures \ref{fig:NuRe}(a,b,c) show the normalized $Nu$ against $1/\Gamma$ for CRB, $1/Ro$ for RRB, and $\Lambda$ for DDC, respectively. It is clearly seen that moderate stabilization for all systems can enhance the global heat or salinity transport. Given that the flow becomes weaker as shown by the reduction of $Re$ in Figs. \ref{fig:NuRe}(d,e,f), the enhancement is non-trivial and counterintuitive. By comparing the three systems side by side, we can conclude that the leading effect of stabilization is similar: Under moderate strength of stabilization, $Nu$ first increases with increasing stabilizing parameters; however, excessive stabilization will eventually lead to the sharp decline in $Nu$. We note that each of the three cases might be different in some details. For example, there is a pronounced optimal point in CRB and RRB at which $Nu$ attains a maximum value but the optimal state in DDC is achieved over a range of density ratio $\Lambda$ instead of at a single point. Nonetheless, from the similarity recognized here we expect that there might be some unifying mechanisms for the systems under distinct forms of stabilizing forces and a more fundamental understanding on this class of stabilized turbulent flows might emerge.

\begin{figure*}
\includegraphics[width=0.95\textwidth]{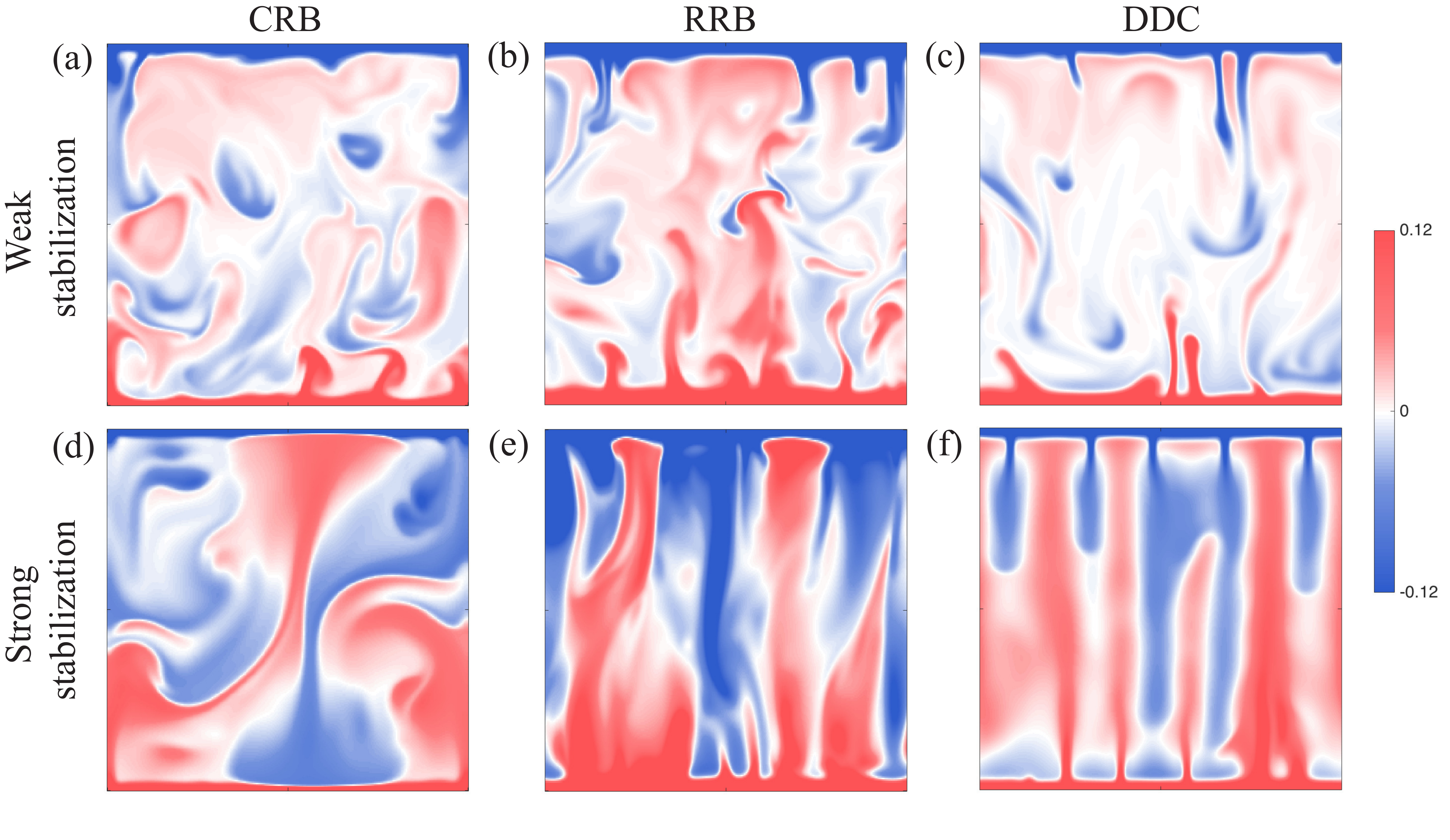}
\caption{\label{fig:scalarfield}Instantaneous scalar fields for CRB, RRB and DDC with all plots fixed at $Ra=1\times10^8$ at (a) $1/\Gamma=2$, (b) $1/Ro=1$, (c) $\Lambda=0.01$,  (d) $1/\Gamma=10$, (e) $1/Ro=7$ and (f) $\Lambda=4$. The scalar fields are taken at the middle vertical plane and it is the midway along the confinement direction in CRB. Here the reddish (bluish) color represents the hot (cold) fluid in CRB and RRB, while reddish and bluish colors represent the fresher and saltier fluid in DDC. Note that in RRB and DDC, only part of the periodic domain is shown here.}
\end{figure*}

In this paper we will work out this universality. In Fig.~\ref{fig:scalarfield} we compare the flow morphologies at the middle vertical plane (in CRB the midway along the confinement direction) with weak and moderate stabilization for the three systems ($Ra=1\times10^8$, salinity-$Ra$ number for DDC). First, the thermal (or salinity) structures are shown in figure \ref{fig:scalarfield} (a,b,c) at $1/\Gamma=2$, $1/Ro=1$ and $\Lambda=0.01$ which correspond to the state with moderate scalar transport enhancement and thus the effects of confinement, rotation, and temperature stabilization may be considered to be weak in the respective systems. As seen from~\ref{fig:scalarfield} (a,b), for CRB and RRB large portions of heat are carried by the mushroom-like plumes which are detaching from the top and bottom boundary layers. When the thermal plumes propagate vertically, their heat content diffuses to the turbulent bulk progressively and their coherency is lost when reaching the opposite boundary layers. In DDC, the salinity structures appear to be more slender than the thermal structures in CRB and RRB because of the large salinity Prandtl number. It is clear from the above observation that under very weak stabilization forces the morphologies of the  thermal and salinity structures are similar to that in classical Rayleigh-B\'enard flow. 

\begin{figure*}
\includegraphics[width=0.95\textwidth]{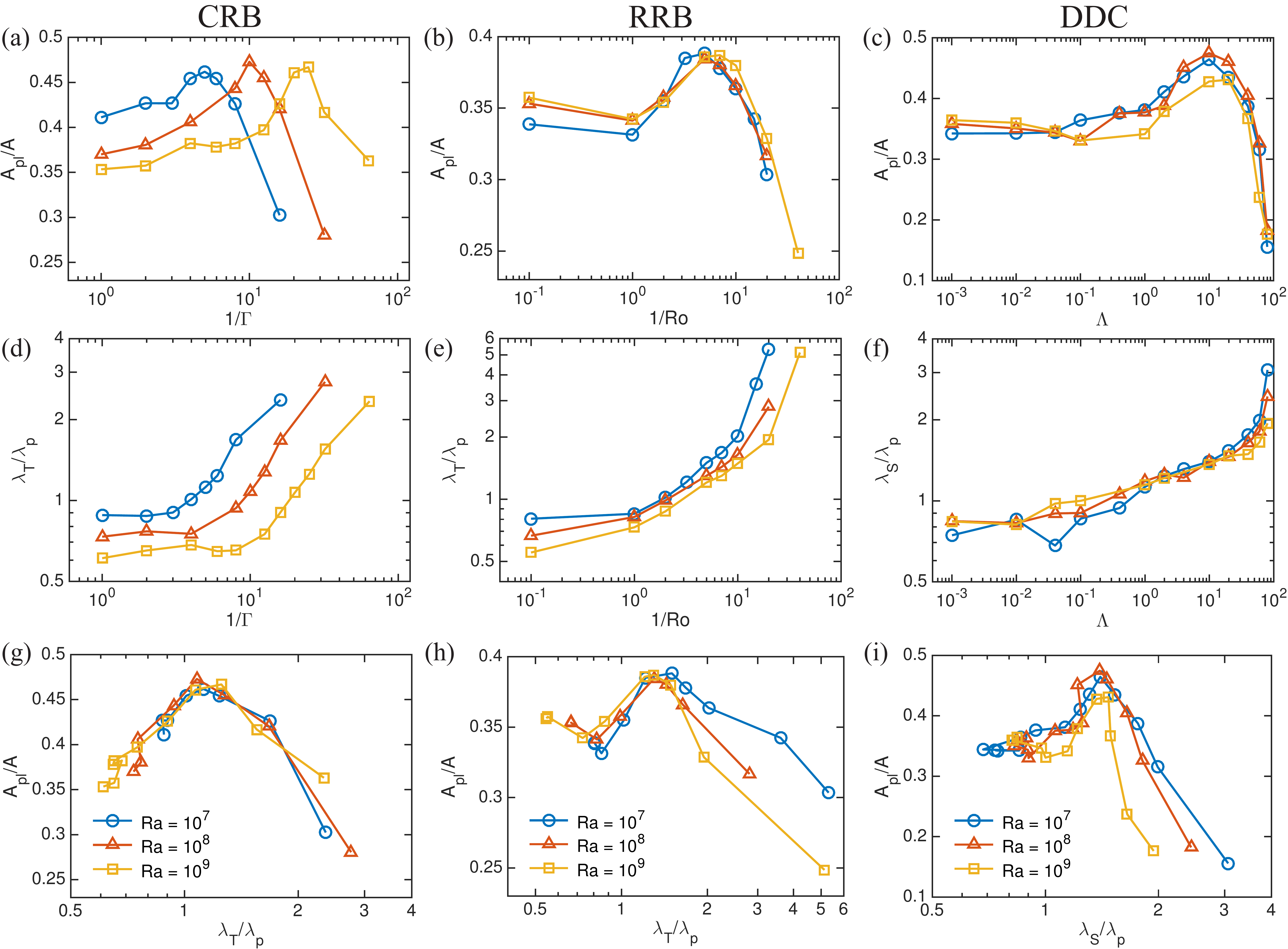}
\caption{\label{fig:bdcrossing}Plume coverage $A_{pl}/A$ evaluated at the edge of thermal/salinity boundary layer versus (a) $1/\Gamma$, (b) $1/Ro$ and (c) $\Lambda$. Here $A_{pl}$ is the area covered by cold/saline fluid and $A$ is the total area. Ratio of the thermal/salinity boundary layer thickness over the momentum one versus (d) $1/\Gamma$, (e) $1/Ro$ and (f) $\Lambda$. Plume coverage $A_{pl}/A$ versus the relative thickness $\lambda_T/\lambda_p$ for CRB in (g) and RRB in (h), $\lambda_S/\lambda_p$ for DDC in (i).}
\end{figure*}

In contrast to the weakly stabilized cases, the flow morphologies can change considerably under stronger stabilization. Figures \ref{fig:scalarfield} (d,e,f) show the morphologies at $1/\Gamma=10$, $1/Ro=7$ and $\Lambda=4$ which are cases with maximum Nusselt number in the three systems. As the bulk becomes less turbulent by the respective stabilizing forces, highly coherent structures that extend over the entire height of the cell are formed. We remark that the coherent structures in CRB are still wavy at the optimal state. However, under even stronger confinement the system enters into the so-called severely confined regime, with finger-like, long-lived plume columns \citep{chong2016jfm} similar to those observed in RRB and DDC. By examining the plume coverage we show below that  this formation of highly coherent thermal/salinity plumes is crucial to the enhanced scalar transport, since more coherent structures can better preserve their heat/salt content against thermal/molecular diffusion when traversing to the opposite boundary layer. Figures \ref{fig:bdcrossing} (a,b,c) show the portion of area covered by the cold/salty fluid $A_{pl}/A$ at the edge of the bottom thermal/salinity boundary layer. Indeed, moderate strength of stabilization can lead to larger portions of cold/salty plumes covering the bottom plate as compared to the weakly stabilized cases. It also shows that a too strong stabilization can eventually cause the rapid drop in plume coverage, which coincides with the decline of the global heat/salinity transport.

We have thus revealed that a hallmark of the stabilizing-destabilizing  (S-D) turbulent flow is the formation of highly coherent structures. These structures can extend over the height of the cell. In all three systems, the plumes grow from the boundary layer regions that carry the high temperature/salinity anomaly. So we now turn to the effects of the stabilizing mechanism on the boundary layer behaviors. Figures \ref{fig:bdcrossing} (d,e,f) show the ratio  ($\lambda_T/\lambda_p$ or $\lambda_S/\lambda_p$) of the thermal/salinity boundary layer thickness over the momentum boundary layer thickness, where the thermal (or salinity) BL thickness $\lambda_T$ (or $\lambda_S$) is defined by the first peak of the temperature (or salinity) standard deviation profile from the bottom and the momentum BL $\lambda_p$ is defined by the position of the first peak of $(\partial_x u)^2+(\partial_y v)^2+(\partial_z w)^2$ profile, i.e. the location with maximum stress. The figures show that the boundary layer thickness ratios increase with the increase of stabilization forces and eventually the momentum boundary layer becomes thinner than the thermal (or salinity) boundary layer. Note that the momentum boundary layer defined by the stress is different from the traditional definition using the first peak of horizontal velocity root-mean-square because the maximum horizontal velocity does not necessarily lead to the maximum stress (see Supplementary Materials for the details). We have found that the boundary layer thickness defined by the stresses is more relevant to the mechanism involved here than the traditional definition. Furthermore, we plot the plume coverage versus the ratio of boundary layer thickness $\lambda_T$/$\lambda_p$ or $\lambda_S$/$\lambda_p$ in figures \ref{fig:bdcrossing} (g,h,i). It is clear that the plume coverage reaches maximum (corresponding to maximum transport enhancement) when the thickness ratio becomes larger than a certain value of order unity and then declines sharply afterwards. This suggests that the necessary condition for the global transport to keep increasing is that the momentum boundary layer not becoming too thin comparing to the thermal/salinity BL; otherwise the hot/fresh fluid cannot be well-accumulated or well-organized before being ejected at the edge of momentum boundary layer, and further suppression on the strength of bulk flow leads to the sharp decline in scalar transport efficiency. This strong correlation between plume coverage and boundary crossing shared by all three systems is another salient feature of S-D flow.

In summary, we have investigated RB flow with a stabilization force using three examples, i.e. the viscous, Corioslis, and negative buoyancy forces. For all three flows we observed significant transport enhancement for moderate values of the stabilizing force. Despite the fact that the nature of the three stabilizing forces are very different, our analysis shows that these forces can similarly influence the coherent structures and the boundary layers. Our study therefore reveals a universal mechanism underpinning scalar transport enhancement in the three types of stabilizing-destabilizing (S-D) flows.
For an appropriate strength of the stabilizing force, the flow structures become more coherent with the vertical motions severely suppressed, resulting in a higher efficiency of scalar transport. We stress that this class of flow might be generalized to other situations involving different stabilizing forces, such as the Lorentz force in convection with conducting fluid under vertical magnetic field. Indeed, the role of Lorentz force in stabilizing turbulent convection is known~\citep{chandrasekhar2013hydrodynamic,aurnou2001jfm} leading to a larger critical Rayleigh number with stronger external magnetic field. The ability to understand similar phenomena occurring in different systems under a unified framework has been a hallmark of physics research. The present study of stabilizing-destabilizing flows provide one such example and may therefore inspire work on other systems.

\bigskip

K.L.C. and Y.Y. contributed equally to this work. K.L.C., S.D.H., and K.Q.X. were supported in part by the Hong Kong Research Grants Council under Grant No. CUHK404513 and a NSFC$/$RGC Joint Research Grant N\_CUHK437$/$15, and through a Hong Kong PhD Fellowship. J.Q.Z. was partially supported by a NSFC$/$RGC Joint Research Grant No.~11561161004. Y.Y., R.J.A.M.S., R.V., and D.L. acknowledge the support from the Dutch Foundation for Fundamental Research on Matter (FOM), and by the Netherlands Center for Multiscale Catalytic Energy Conversion (MCEC), an NWO Gravitation programme funded by the Ministry of Education, Culture and Science of the government of the Netherlands. The computing resources for CRB were provided by the Leibnitz-Rechenzentrum Munich under the Project No. pr47vi and the High Performance Cluster Computing Centre, Hong Kong Baptist University, and those for RRB and DDC were provided by the Dutch national e-infrastructure of SURFsara, and the Marconi supercomputer based in CINECA, Italy through the PRACE project number 2016143351.


\begin{thebibliography}{10}

\bibitem{ahlers2009rmp}
G.~Ahlers, S.~Grossmann, and D.~Lohse,
\newblock Rev. Mod. Phys. {\bf 81}, 503 (2009).

\bibitem{lohse2010arfm}
D.~Lohse and K.-Q. Xia,
\newblock Annu. Rev. Fluid Mech. {\bf 42}, 335 (2010).

\bibitem{chilla2012epj}
F.~Chill\`a and J.~Schumacher,
\newblock Eur. Phys. J. E {\bf 35}, 58 (2012).

\bibitem{xia2013taml}
K.-Q. Xia,
\newblock Theor. Appl. Mech. Lett. {\bf 3}, 052001 (2013).

\bibitem{proudman1916prsla}
J.~Proudman,
\newblock Proc. R. Soc. Lond. A {\bf 92}, 408 (1916).

\bibitem{taylor1917prsla}
G.~I. Taylor,
\newblock Proc. R. Soc. Lond. A {\bf 93}, 92 (1917).

\bibitem{chandrasekhar2013hydrodynamic}
S.~Chandrasekhar,
\newblock {\em Hydrodynamic and hydromagnetic stability} (Clarendon Press,
  Oxford, 1961).

\bibitem{radko2013double}
T.~Radko,
\newblock {\em Double-diffusive convection} (Cambridge {U}niversity {P}ress,
  2013).

\bibitem{busse1994chaos}
F.~H. Busse,
\newblock Chaos {\bf 4}, 123 (1994).

\bibitem{merryfield1995aj}
W.~J. Merryfield,
\newblock Astrophys. J. {\bf 444}, 318 (1995).

\bibitem{rosenblum2011aj}
E.~Rosenblum, P.~Garaud, A.~Traxler, and S.~Stellmach,
\newblock Astrophys. J. {\bf 731}, 66 (2011).

\bibitem{leconte2012aa}
J.~Leconte and G.~Chabrier,
\newblock Astron. Astrophys. {\bf 540}, A20 (2012).

\bibitem{mirouh2012aj}
G.~M. Mirouh, P.~Garaud, S.~Stellmach, A.~L. Traxler, and T.~S. Wood,
\newblock Astrophys. J. {\bf 750}, 61 (2012).

\bibitem{wood2013aj}
T.~S. Wood, P.~Garaud, and S.~Stellmach,
\newblock Astrophys. J. {\bf 768}, 157 (2013).

\bibitem{schoofs1999epsl}
S.~Schoofs, F.~J. Spera, and U.~Hansen,
\newblock Earth Planet Sci. Lett. {\bf 174}, 213 (1999).

\bibitem{buffett2010jgrse}
B.~A. Buffett and C.~T. Seagle,
\newblock J. Geophys. Res. Solid Earth {\bf 115}, B04407 (2010).

\bibitem{turner1985arfm}
J.~Turner,
\newblock Annu. Rev. Fluid Mech. {\bf 17}, 11 (1985).

\bibitem{schmitt1994arfm}
R.~W. Schmitt,
\newblock Annu. Rev. Fluid Mech. {\bf 26}, 255 (1994).

\bibitem{marshall1999rg}
J.~Marshall and F.~Schott,
\newblock Rev. Geophys. {\bf 37}, 1 (1999).

\bibitem{schmitt2005science}
R.~W. Schmitt, J.~R. Ledwell, E.~T. Montgomery, K.~L. Polzin, and J.~M. Toole,
\newblock Science {\bf 308}, 685 (2005).

\bibitem{nada2007ijhmt}
S.~A. Nada,
\newblock Int. J. Heat Mass Transfer {\bf 50}, 667 (2007).

\bibitem{huang2013prl}
S.-D. Huang, M.~Kaczorowski, R.~Ni, and K.-Q. Xia,
\newblock Phys. Rev. Lett. {\bf 111}, 104501 (2013).

\bibitem{wagner2013pof}
S.~Wagner and O.~Shishkina,
\newblock Phys. Fluids {\bf 25}, 1 (2013).

\bibitem{chong2015prl}
K.~L. Chong, S.-D. Huang, M.~Kaczorowski, and K.-Q. Xia,
\newblock Phys. Rev. Lett. {\bf 115}, 264503 (2015).

\bibitem{king2009nature}
E.~M. King, S.~Stellmach, J.~Noir, U.~Hansen, and J.~M. Aurnou,
\newblock Nature {\bf 457}, 301 (2009).

\bibitem{zhong2009prl}
J.~Q. Zhong, R.~J. A.~M. Stevens, H.~J.~H. Clercx, R.~Verzicco, D.~Lohse and G.~Ahlers,
\newblock Phys. Rev. Lett. {\bf 102}, 044502 (2009).

\bibitem{stevens2009prl}
R.~J. A.~M. Stevens, J.~Q. Zhong, H.~J.~H. Clercx, G.~Ahlers, and D.~Lohse,
\newblock Phys. Rev. Lett. {\bf 103}, 024503 (2009).

\bibitem{wei2015prl}
P.~Wei, S.~Weiss, and G.~Ahlers,
\newblock Phys. Rev. Lett. {\bf 114}, 114506 (2015).

\bibitem{kellner2014pof}
M.~Kellner and A.~Tilgner,
\newblock Phys. Fluids {\bf 26}, 094103 (2014).

\bibitem{yang2016pnas}
Y.~Yang, R.~Verzicco, and D.~Lohse,
\newblock PNAS {\bf 113}, 69 (2016).

\bibitem{kaczorowski2008nrnefm6}
M.~Kaczorowski, A.~Shishkin, O.~Shishkina, and C.~Wagner,
\newblock New Results in Numerical and Experimental Fluid Mech. VI {\bf 96},
  381 (2008).

\bibitem{kaczorowski2013jfm}
M.~Kaczorowski and K.-Q. Xia,
\newblock J. Fluid Mech. {\bf 722}, 596 (2013).

\bibitem{kaczorowski2014jfm}
M.~Kaczorowski, K.-L. Chong, and K.-Q. Xia,
\newblock J. Fluid Mech. {\bf 747}, 73 (2014).

\bibitem{yang2015jfm}
Y.~Yang, E.~P. van~der Poel, R.~Ostilla-{M}onico, S.~Chao, R.~Verzicco, S.~Grossmann and D.~Lohse,
\newblock J. Fluid Mech. {\bf 768}, 476 (2015).

\bibitem{yang2016jfm}
Y.~Yang, R.~Verzicco, and D.~Lohse,
\newblock J. Fluid Mech. {\bf 802}, 667 (2016).

\bibitem{ostillamonico2015jcp}
R.~Ostilla-{M}onico, Y.~Yang, E.~P. van~der Poel, D.~Lohse, and R.~Verzicco,
\newblock J. Comput. Phys. {\bf 301}, 308–321 (2015).

\bibitem{chong2016jfm}
K.~L. Chong and K.-Q. Xia,
\newblock J. Fluid Mech. {\bf 805}, R4 (2016).

\bibitem{aurnou2001jfm}
J.~M. Aurnou and P.~L. Olson,
\newblock J. Fluid Mech. {\bf 430}, 283 (2001).

\end{thebibliography}

\newpage

\section{Supplementary Materials}

In this Supplementary Materials, we discuss how to evaluate the coverage of thermal/salinity plumes and the thickness of momentum boundary layer. In addition, we provide the numerical details and results of all simulation cases in table \ref{tab:table1} for CRB, table \ref{tab:table2} for RRB and table \ref{tab:table3} for DDC.

We have examined the portion of area covered by the cold/salty plumes at the edge of the bottom thermal/salinity boundary layer. To estimate the coverage, we first extract the cold/salty plumes by the fact that they carry high temperature/salinity anomaly. From a single snapshot, we calculate the area satisfying $-(T-\langle T \rangle_{xy}) \geq cT_{rms}$ (or $S-\langle S \rangle_{xy} \geq cS_{rms}$ for DDC). The rms value of the non-stabilized counterpart ($1/\Gamma=1$, $1/Ro=0$ or $\Lambda=0$) is used for the same $Ra$ instead of their individual rms, such that the threshold is the same for the series of data with the same $Ra$. The empirical parameter $c$ is chosen to be $0.5$, and the subscript xy denotes averaging over the horizontal slice. We have tried a few values of $c$ and found that the qualitative feature of the result is not sensitive to the value of $c$. We calculate the coverage for successive snapshots separated by several free-fall time intervals, and the mean coverage can be estimated by their ensemble average.
\vspace*{-8mm}
\begin{figure}[h]
\includegraphics[width=1.0\textwidth]{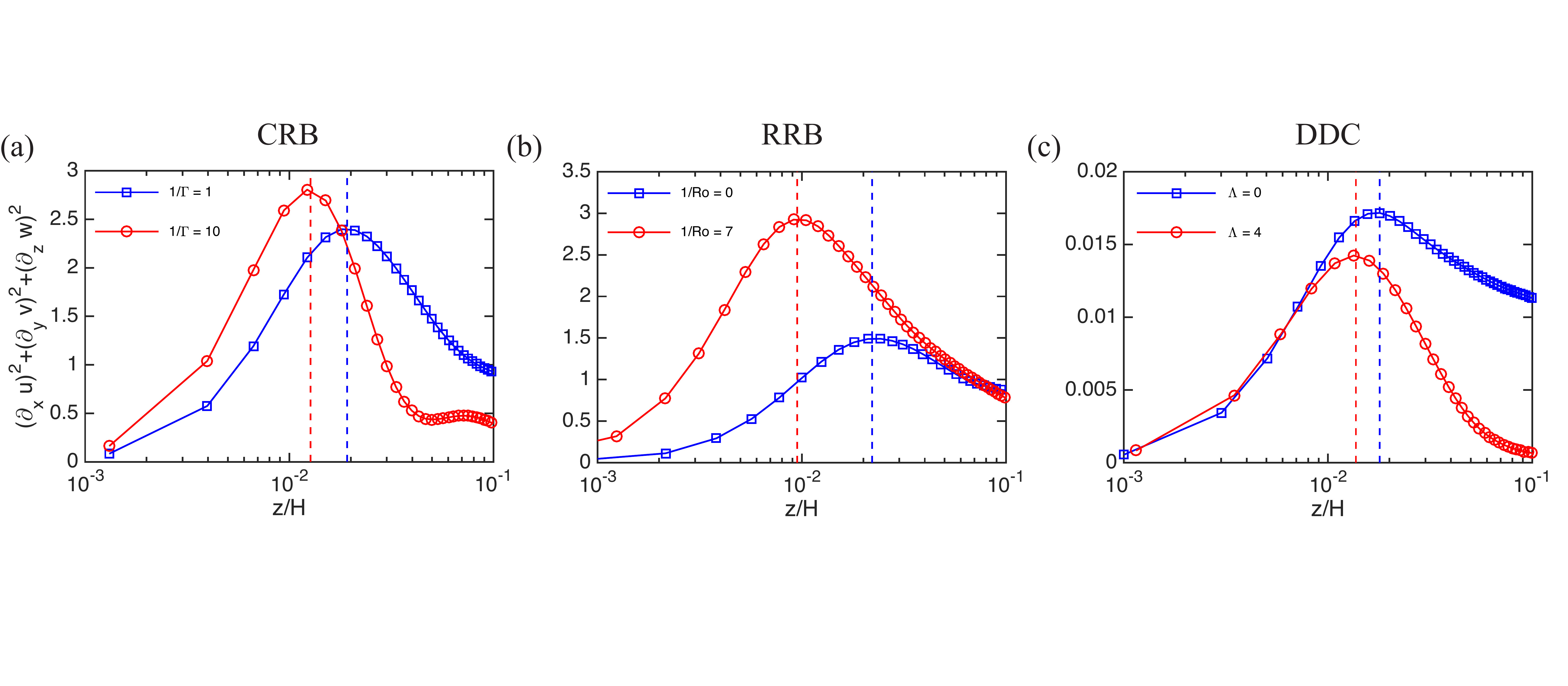}
\vspace*{-20mm}
\caption{\label{fig:stressprof}Vertical profiles of $(\partial_x u)^2+(\partial_y v)^2+(\partial_z w)^2$ averaged horizontally for CRB in (a), RRB in (b) and DDC in (c). The vertical dashed line indicates the location of peak for each profile and this peak is obtained by a quadratic fitting from the three nearby points.}
\end{figure}

Example vertical profiles of the quantity $(\partial_x u)^2+(\partial_y v)^2+(\partial_z w)^2$ have been plotted in Figs. S1 (a,b,c) for CRB, RRB and DDC, respectively. The quantity is the square of the normal gradient of velocity summing over all components, which measures the overall magnitude of the normal stress. From the figures, each stress profile shows a well-defined peak at the location very close to the bottom plate which we consider to be the location with strongest plume merging and convolution take place. We define the momentum boundary layer thickness by the position of this peak from the stress profile. To better estimate this location, we first select the three nearby points around this peak and then adopt a quadratic fitting. We also compare the profile of the stress and the rms of the horizontal velocity in Figs. S2 (a,b). The figure shows that the peak location of the maximum stress is different from that of the maximum rms of horizontal velocity. As the quantity $(\partial_x u)^2+(\partial_y v)^2+(\partial_z w)^2$ measures the strength of upward fluid suction of the hot/fresh fluid near the bottom plate, which is directly related to heat/salt transfer, we use the peak position of this quantity as the momentum boundary layer in the present paper.
\vspace*{-8mm}
\begin{figure}[h]
\includegraphics[width=0.8\textwidth]{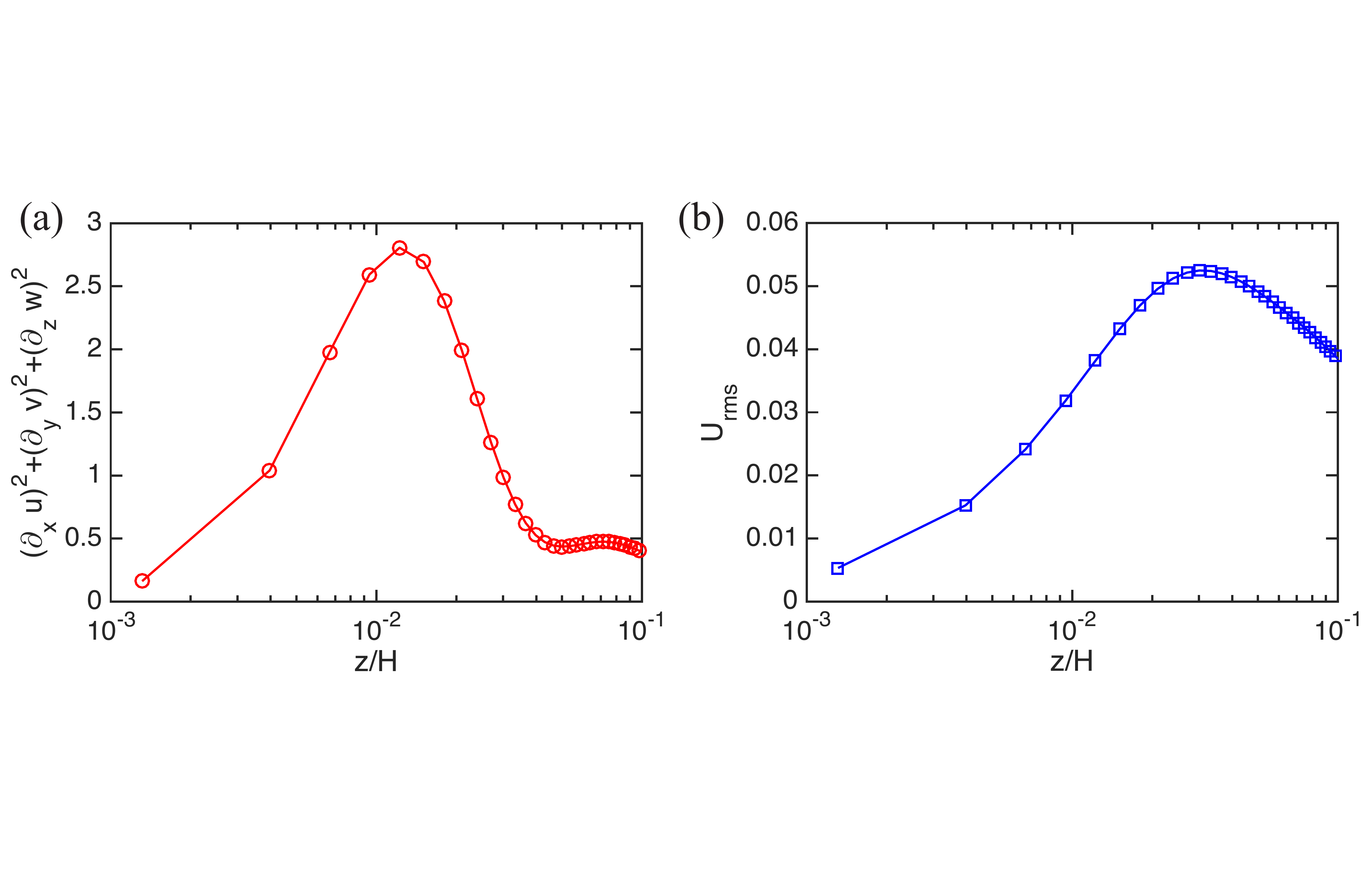}
\vspace*{-20mm}
\caption{\label{fig:compareprof}Vertical profiles of $(\partial_x u)^2+(\partial_y v)^2+(\partial_z w)^2$ in (a) and the root-mean-square of horizontal velocity $u_{rms}$ in (b) for $1/\Gamma=10$ in CRB.}
\end{figure}

\begin{table}[b!]
\begin{ruledtabular}
\begin{tabular}{cccccccc}
$Ra$&$1/\Gamma$&$Nu$&$Re$&$\lambda_T$&$\lambda_p$ & $A_{pl}/A$ &$N_x \times N_y \times N_z$\\
\hline
$10^7$ 	& 1  	& 16.1 & 129.1 & 2.90E-2	&  3.29E-2 & 0.411  & $176 \times 176 \times 192$\\
$10^7$ 	& 2 	& 15.9 & 115.3 & 2.96E-2	&  3.38E-2 & 0.427  & $176 \times 88 \times 192$ \\
$10^7$ 	& 3   	& 16.2 & 106.6 & 2.87E-2	&  3.19E-2 & 0.427  & $176 \times 64 \times 192$ \\
$10^7$ 	& 4   	& 16.7 & 103.1 & 2.83E-2	&  2.80E-2 & 0.454  & $176 \times 48 \times 192$ \\
$10^7$ 	& 5   	& 16.8 & 97.5  & 2.81E-2	&  2.51E-2 & 0.462  & $176 \times 40 \times 192$ \\
$10^7$ 	& 6   	& 16.6 & 92.6  & 2.90E-2	&  2.34E-2 & 0.454  & $176 \times 36 \times 192$ \\
$10^7$ 	& 8   	& 14.8 & 76.5  & 3.50E-2	&  2.08E-2 & 0.426  & $176 \times 24 \times 192$ \\
$10^7$ 	& 16 	& 9.5  & 37.6  & 7.07E-2	&  2.99E-2 & 0.302  & $176 \times 16 \times 192$ \\ 

$10^8$ & 1  	& 32.0 & 458.0  &1.41E-2   &1.92E-2   &0.370   & $256 \times 256 \times 256$  \\
$10^8$ & 2  	& 31.7 & 391.9  &1.49E-2   &1.94E-2   &0.381   & $256 \times   134 \times 256$ \\
$10^8$ & 4  	& 32.2 & 335.7  &1.42E-2   &1.89E-2   &0.406   & $256 \times 72 \times 256$    \\
$10^8$ & 8  	& 34.6 & 313.8  &1.40E-2   &1.49E-2   &0.443   & $256 \times 38 \times 256$    \\
$10^8$ & 10 	& 35.4 & 289.6  &1.37E-2   &1.27E-2   &0.473   & $256 \times 36 \times 256$    \\
$10^8$ & 12.5 	& 34.9 & 251.0  &1.35E-2   &1.06E-2   &0.455   & $256 \times 32 \times 256$    \\
$10^8$ & 16 	& 29.0 & 194.2  &1.56E-2   &9.35E-3   &0.420   & $256 \times 24 \times 256$ 	 \\
$10^8$ & 32 	& 19.7 & 92.5   &3.47E-2   &1.26E-2   &0.280   & $256 \times 20 \times 256$ 	 \\

$10^9$ & 1 			& 62.7 & 1500.5  &  6.88E-3  & 1.13E-2  & 0.353 & $512 \times 512 \times 512$ 	   \\
$10^9$ & 2 			& 63.2 & 1363.3  &  6.97E-3  & 1.07E-2  & 0.358 & $512 \times   274 \times 512$    \\
$10^9$ & 4 			& 63.6 & 1147.5  &  7.18E-3  & 1.05E-2  & 0.382 & $512 \times 154 \times 512$      \\
$10^9$ & 6 			& 65.3 & 1064.7  &  6.80E-3  & 1.05E-2  & 0.378 & $512 \times 128 \times 512$      \\
$10^9$ & 8 			& 67.8 & 1029.1  &  6.62E-3  & 1.01E-2  & 0.382 & $512 \times 96 \times 512$   	   \\
$10^9$ & 12.5 		& 70.6 & 925.0   &  6.48E-3  & 8.64E-3  & 0.397 & $512 \times 80 \times 512$   	   \\
$10^9$ & 16 		& 72.6 & 855.1   &  6.63E-3  & 7.33E-3  & 0.426 & $512 \times 64 \times 512$   	   \\
$10^9$ & 20 		& 74.1 & 770.1   &  6.59E-3  & 6.16E-3  & 0.461 & $512 \times 56 \times 512$ 	   \\
$10^9$ & 25 		& 72.3 & 653.4   &  6.44E-3  & 5.14E-3  & 0.467 & $512 \times 48 \times 512$ 	   \\
$10^9$ & 32 		& 61.3 & 494.8   &  6.65E-3  & 4.28E-3  & 0.417 & $512 \times 32 \times 512$ 	   \\
$10^9$ & 64 		& 47.0 & 235.1   &  1.06E-2  & 4.55E-3  & 0.363 & $512 \times 28 \times 512$ 	   \\
\end{tabular}
\end{ruledtabular}
\caption{\label{tab:table1}Simulation control parameters and numerical results in CRB. All simulations at are $Pr=4.38$. Columns from left to right: the Rayleigh number $Ra$, the reciprocal of the width-to-height ratio $1/\Gamma$, the Nusselt number $Nu$, the Reynolds number $Re$, the thickness of temperature and momentum boundary layers $\lambda_T$ and $\lambda_p$, the plume coverage $A_{pl}/A$ and the grid resolutions along x, y and z directions. }
\end{table}

\begin{table}[h!]
\begin{ruledtabular}
\begin{tabular}{cccccccccc}
$Ra$&$1/Ro$&$\Gamma$&$Nu$&$Re$&$\lambda_T$&$\lambda_p$ & $A_{pl}/A$ &$N_x \times N_z$&$n_x \times n_z$\\
\hline
$10^7$ & 0.0   &5   &16.3  &143.3  &2.92E-2   &2.92E-2  &0.338   &$240 \times 144$   &$2 \times 1$ \\  
$10^7$ & 0.1   &5   &16.3  &143.7  &2.92E-2   &2.92E-2  &0.339   &$240 \times 144$   &$2 \times 1$ \\  
$10^7$ & 1.0   &5   &18.6  &104.5  &2.59E-2   &2.59E-2  &0.331   &$240 \times 144$   &$2 \times 1$ \\  
$10^7$ & 2.0   &5   &19.5  &96.2   &2.69E-2   &2.69E-2  &0.355   &$288 \times 144$   &$2 \times 1$ \\  
$10^7$ & 3.2   &5   &20.1  &89.6   &2.77E-2   &2.77E-2  &0.385   &$288 \times 144$   &$2 \times 1$ \\  
$10^7$ & 5.0   &4   &20.4  &82.3   &2.82E-2   &2.82E-2  &0.388   &$240 \times 144$   &$2 \times 1$ \\  
$10^7$ & 7.0   &4   &20.0  &75.3   &2.83E-2   &2.83E-2  &0.378   &$240 \times 144$   &$2 \times 1$ \\  
$10^7$ & 10.0  &4   &17.8  &65.0   &3.09E-2   &3.09E-2  &0.364   &$240 \times 144$   &$2 \times 1$ \\  
$10^7$ & 15.0  &4   &12.5  &48.1   &4.16E-2   &4.16E-2  &0.342   &$240 \times 144$   &$2 \times 1$ \\  
$10^7$ & 20.0  &4   &7.4   &33.0   &5.83E-2   &5.83E-2  &0.303   &$240 \times 192$   &$2 \times 1$ \\  

$10^8$ & 0.0   &4   &31.0  &480.0  &1.47E-2   &2.21E-2  &0.353   &$288 \times 144$   &$3 \times 2$ \\
$10^8$ & 0.1   &4   &31.3  &459.2  &1.47E-2   &2.20E-2  &0.353   &$288 \times 144$   &$3 \times 2$ \\
$10^8$ & 1.0   &3   &34.7  &299.9  &1.35E-2   &1.64E-2  &0.341   &$288 \times 192$   &$3 \times 2$ \\
$10^8$ & 2.0   &3   &36.9  &276.9  &1.34E-2   &1.36E-2  &0.358   &$288 \times 192$   &$3 \times 2$ \\
$10^8$ & 5.0   &3   &39.8  &243.2  &1.36E-2   &1.05E-2  &0.385   &$288 \times 192$   &$3 \times 2$ \\
$10^8$ & 7.0   &3   &39.9  &225.7  &1.36E-2   &9.50E-3  &0.380   &$360 \times 240$   &$3 \times 1$ \\
$10^8$ & 10.0  &3   &38.8  &205.3  &1.34E-2   &8.17E-3  &0.366   &$360 \times 240$   &$3 \times 1$ \\
$10^8$ & 20.0  &2   &25.6  &137.6  &1.69E-2   &6.01E-3  &0.317   &$240 \times 240$   &$3 \times 1$ \\

$10^9$ & 0.0   &3   &61.7  &1495.0 &7.19E-3   &1.32E-2  &0.356   &$384 \times 192$   &$4 \times 2$ \\
$10^9$ & 0.1   &3   &62.4  &1414.0 &7.15E-3   &1.29E-2  &0.357   &$384 \times 192$   &$4 \times 2$ \\
$10^9$ & 1.0   &1   &66.5  &865.0  &6.62E-3   &9.05E-3  &0.342   &$288 \times 240$   &$2 \times 2$ \\
$10^9$ & 2.0   &1   &68.6  &764.7  &6.63E-3   &7.59E-3  &0.354   &$360 \times 240$   &$2 \times 2$ \\
$10^9$ & 5.0   &1   &71.4  &637.3  &6.79E-3   &5.63E-3  &0.386   &$360 \times 240$   &$2 \times 2$ \\
$10^9$ & 7.0   &1   &72.5  &608.4  &6.77E-3   &5.22E-3  &0.387   &$360 \times 240$   &$2 \times 2$ \\
$10^9$ & 10.0  &1   &71.7  &562.2  &6.61E-3   &4.44E-3  &0.380   &$360 \times 240$   &$2 \times 2$ \\
$10^9$ & 20.0  &1   &63.0  &461.8  &6.45E-3   &3.32E-3  &0.329   &$384 \times 240$   &$2 \times 2$ \\
$10^9$ & 40.0  &1   &27.5  &249.4  &1.25E-2   &2.44E-3  &0.249   &$288 \times 288$   &$3 \times 1$ \\
\end{tabular}
\end{ruledtabular}
\caption{\label{tab:table2}Simulation control parameters and numerical results in RRB. All simulations at are $Pr=6.4$ with periodic boundary condition. Columns from left to right: the Rayleigh number $Ra$, the reciprocal Rossby number $1/Ro$, the aspect-ratio of the computational domain $\Gamma$, the Nusselt number $Nu$, the Reynolds number $Re$, the thickness of temperature and momentum boundary layers $\lambda_T$ and $\lambda_p$, the plume coverage $A_{pl}/A$ and the base grid resolutions and refinement factors along x and z directions for the temperature field. Note that the resolutions in y direction is exactly the same as those in x direction.}
\end{table}

\begin{table}[h!]
\begin{ruledtabular}
\begin{tabular}{cccccccccc}
$Ra_S$&$\Lambda$&$\Gamma$&$Nu_S$&$Re$&$\lambda_S$&$\lambda_p$ & $A_{pl}/A$ &$N_x \times N_z$&$n_x \times n_z$\\
\hline
$10^7$ &   0.0E+0   &2.0   &17.3	&1.680	&3.12E-2	&4.25E-2  &0.343   &$192 \times 144$  &$3 \times 2$ \\
$10^7$ &   1.0E-3   &2.0   &17.3	&1.666	&3.12E-2	&4.19E-2  &0.342   &$192 \times 144$  &$3 \times 2$ \\
$10^7$ &   1.0E-2   &2.0   &17.7	&1.446	&3.04E-2	&3.57E-2  &0.343   &$192 \times 144$  &$3 \times 2$ \\
$10^7$ &   4.0E-2   &2.0   &18.0	&1.244	&3.01E-2	&4.40E-2  &0.344   &$192 \times 144$  &$3 \times 2$ \\
$10^7$ &   1.0E-1   &2.0   &18.4	&1.069	&2.99E-2	&3.48E-2  &0.364   &$192 \times 144$  &$3 \times 2$ \\
$10^7$ &   4.0E-1   &2.0   &18.8	&0.830	&3.03E-2	&3.22E-2  &0.376   &$192 \times 120$  &$3 \times 2$ \\
$10^7$ &   1.0E+0   &2.0   &18.3	&0.671	&3.18E-2	&2.82E-2  &0.381   &$160 \times 120$  &$3 \times 2$ \\
$10^7$ &   2.0E+0   &2.0   &18.3	&0.577	&3.22E-2	&2.60E-2  &0.411   &$144 \times 120$  &$3 \times 2$ \\
$10^7$ &   4.0E+0   &2.0   &18.1	&0.482	&3.31E-2	&2.52E-2  &0.436   &$144 \times 120$  &$3 \times 2$ \\
$10^7$ &   1.0E+1   &2.0   &17.4	&0.359	&3.36E-2	&2.40E-2  &0.465   &$192 \times 144$  &$2 \times 2$ \\
$10^7$ &   2.0E+1   &2.0   &15.3	&0.281	&3.45E-2	&2.25E-2  &0.434   &$120 \times 120$  &$3 \times 2$ \\
$10^7$ &   4.0E+1   &2.0   &10.7	&0.176	&3.97E-2	&2.25E-2  &0.388   &$144 \times 96$  &$2 \times 2$ \\
$10^7$ &   6.0E+1   &2.0   &6.1	&0.112	&4.73E-2	&2.38E-2  &0.315   &$144 \times 96$  &$2 \times 2$ \\
$10^7$ &   8.0E+1   &2.0   &2.4	&0.052	&8.14E-2	&2.65E-2  &0.156   &$144 \times 96$  &$2 \times 2$ \\

$10^8$  &   0.0E+0   &1.6    &33.0	&6.217	&1.59E-2	&1.79E-2  &0.363   &$288 \times 288$  &$3 \times 2$ \\
$10^8$  &   1.0E-3   &1.6    &33.1	&5.770	&1.58E-2	&1.88E-2  &0.358   &$288 \times 288$  &$3 \times 2$ \\
$10^8$  &   1.0E-2   &1.6    &33.9	&4.260	&1.57E-2	&1.90E-2  &0.351   &$288 \times 240$  &$3 \times 2$ \\
$10^8$  &   4.0E-2   &1.6    &34.5	&3.537	&1.57E-2	&1.75E-2  &0.344   &$288 \times 240$  &$3 \times 2$ \\
$10^8$  &   1.0E-1   &1.6    &35.5	&2.936	&1.55E-2	&1.72E-2  &0.330   &$288 \times 240$  &$3 \times 2$ \\
$10^8$  &   4.0E-1   &1.6    &36.7	&2.182	&1.58E-2	&1.49E-2  &0.375   &$288 \times 240$  &$3 \times 2$ \\
$10^8$  &   1.0E+0   &1.6    &36.1	&1.741	&1.66E-2	&1.40E-2  &0.377   &$240 \times 240$  &$3 \times 2$ \\
$10^8$  &   2.0E+0   &1.6    &36.1	&1.459	&1.68E-2	&1.34E-2  &0.388   &$240 \times 216$  &$3 \times 2$ \\
$10^8$  &   4.0E+0   &1.6    &36.5	&1.225	&1.67E-2	&1.37E-2  &0.452   &$240 \times 216$  &$3 \times 2$ \\
$10^8$  &   1.0E+1   &1.6    &34.7	&0.906	&1.68E-2	&1.20E-2  &0.475   &$192 \times 192$  &$3 \times 2$ \\
$10^8$  &   2.0E+1   &1.6    &31.0	&0.680	&1.71E-2	&1.17E-2  &0.461   &$192 \times 192$  &$3 \times 2$ \\
$10^8$  &   4.0E+1   &1.6    &21.5	&0.449	&1.85E-2	&1.12E-2  &0.405   &$192 \times 192$  &$3 \times 2$ \\
$10^8$  &   6.0E+1   &1.6    &12.1	&0.284	&2.16E-2	&1.19E-2  &0.326   &$144 \times 144$  &$3 \times 2$ \\
$10^8$  &   8.0E+1   &1.6    &4.5	&0.136	&3.28E-2	&1.34E-2  &0.182   &$144 \times 120$  &$3 \times 2$ \\

$10^9$  &   0.0E+0   &0.8    &62.7	&25.200	&7.57E-3	&9.22E-3  &0.360   &$288 \times 360$  &$4 \times 3$ \\
$10^9$  &   1.0E-3   &0.8    &63.5	&24.770	&7.52E-3	&8.97E-3  &0.364   &$288 \times 360$  &$4 \times 3$ \\
$10^9$  &   1.0E-2   &0.8    &65.1	&22.620	&7.56E-3	&9.26E-3  &0.360   &$288 \times 288$  &$3 \times 3$ \\
$10^9$  &   4.0E-2   &0.8    &67.7	&10.460	&7.64E-3	&7.82E-3  &0.346   &$288 \times 288$  &$3 \times 3$ \\
$10^9$  &   1.0E-1   &0.8    &68.4	&7.752	&7.92E-3	&7.92E-3  &0.331   &$288 \times 288$  &$3 \times 3$ \\
$10^9$  &   1.0E+0   &0.8    &71.1	&4.438	&8.35E-3	&7.29E-3  &0.341   &$288 \times 360$  &$4 \times 3$ \\
$10^9$  &   2.0E+0   &0.6    &71.9	&3.705	&8.32E-3	&6.84E-3  &0.379   &$288 \times 360$  &$4 \times 3$ \\
$10^9$  &   1.0E+1   &0.8    &70.3	&2.347	&8.27E-3	&6.05E-3  &0.428   &$240 \times 240$  &$3 \times 3$ \\
$10^9$  &   2.0E+1   &0.8    &62.0	&1.736	&8.48E-3	&5.76E-3  &0.431   &$216 \times 216$  &$3 \times 3$ \\
$10^9$  &   4.0E+1   &0.8    &43.4	&1.147	&8.99E-3	&6.03E-3  &0.367   &$192 \times 192$  &$3 \times 3$ \\
$10^9$  &   6.0E+1   &0.8    &24.6	&0.725	&1.01E-2	&6.14E-3  &0.237   &$192 \times 192$  &$2 \times 2$ \\
$10^9$  &   8.0E+1   &0.8    &9.2	&0.368	&1.34E-2	&6.90E-3  &0.176   &$240 \times 240$  &$1 \times 1$ \\

\end{tabular}
\end{ruledtabular}
\caption{\label{tab:table3}Simulation control parameters and numerical results in DDC. All simulations are at $Pr_T=7$ and $Pr_S=700$ with periodic boundary condition. Columns from left to right: the salinity Rayleigh number $Ra_S$, the density ratio $\Lambda$, the aspect-ratio of the computational domain $\Gamma$, the salinity Nusselt number $Nu_S$, the Reynolds number $Re$, the thickness of salinity and momentum boundary layers $\lambda_S$ and $\lambda_p$, the plume coverage $A_{pl}/A$ and the grid resolutions and refinement factors along x and z directions for the scalar fields. Note that the resolutions in y direction is exactly the same as those in x direction.}
\end{table}


\end{document}